%% file: ibvp.tex
\documentclass[prd,aps,showkeys,nofootinbib,amsmath,amssymb]{revtex4}
\usepackage{graphicx,color}
\usepackage{amssymb,epsfig}

\newcommand{\Real}{\mathbb{R}}

\newcommand{\eps}{\varepsilon}

\newtheorem{definition}{Definition}
\newtheorem{theorem}{Theorem}
\newtheorem{remark}{Remark}

\newcommand{\qed}{\hfill $\fbox{\hspace{0.3mm}}$ \vspace{.3cm}}

\begin{document}

\title{The Initial-Boundary Value Problem in General Relativity}
\author{Oscar Reula\footnote{FaMAF-UNC, IFEG-CONICET, C\'ordoba, Argentina}}
\affiliation{FaMAF, Universidad Nacional de C\'ordoba, Ciudad Universitaria, 5000, C\'ordoba, Argentina \\ reula@famaf.unc.edu.ar}

\author{Olivier Sarbach}
\affiliation{Instituto de F\'\i sica y Matem\'aticas,
Universidad Michoacana de San Nicol\'as de Hidalgo,\\
Edificio C-3, Ciudad Universitaria, 58040 Morelia, Michoac\'an, M\'exico \\
sarbach@ifm.umich.mx}

\begin{abstract}
In this article we summarize what is known about the initial-boundary value problem for general relativity and discuss present problems related to it.
\end{abstract}


\maketitle

\section{Introduction to the initial-boundary value problem in general relativity}

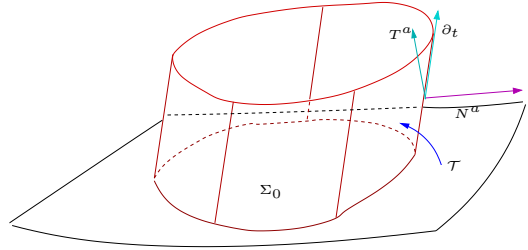
\begin{figure}[ht]
\begin{center}
    \input{prob.pstex_t}
    \caption{The geometrical setup for the initial-boundary value formulation.}
    \label{Fig:IBVP}
\end{center}
\end{figure}

This article has been written to celebrate Mario Castagnino's seventy fifth birthday. Knowing his particular interest in fundamental  problems which has marked his scientific prolific career we are confident he will share our curiosity in this subject.
We consider a manifold $M$ of the form $M = [0,T] \times \Sigma$, where $[0,T]$ is a time interval and $\Sigma$ a three-dimensional manifold with smooth boundaries $\partial\Sigma$, see Fig.~\ref{Fig:IBVP}. We want to answer the following question. What data must be given at the initial surface $\Sigma_0 := \{ 0 \} \times \Sigma$ and the boundary surface ${\cal{T}} := [0,T] \times \partial\Sigma$ such that there exists a unique solution $(M,g)$ of Einstein's vacuum field equations which depends continuously on the data?
From the work of Mme Choquet-Bruhat\cite{cB51} in the fifties we know that at $\Sigma_0$ we must give a pair of symmetric tensor fields $(h_{ab},k_{ab})$, the first being positive definite (and so a metric), which obeys the Einstein constraint equations. They are called the {\sl initial data sets} for Einstein's equations. They give a unique solution to Einstein's equations  on the domain of dependence $D(\Sigma_0)$ of $\Sigma_0$ (which depends on the data given), namely a spacetime $(D(\Sigma_0),g)$, such that $\psi^\star g = h, \psi^\star {\cal L}_n g = k$, where $n$ denotes the unit normal to $\Sigma_0$ in $D(\Sigma_0)$ and $\psi: \Sigma_0 \to D(\Sigma_0)$ the inclusion map. This solution depends continuously on its initial data. This setting is called the {\sl initial value problem}. Furthermore, if two initial data sets are related by a diffeomorphism in $\Sigma_0$, then the unique solutions they produce are also related by a diffeomorphism, this time on the maximal development\cite{cBrG69} of $\Sigma_0$. We call this property the {\sl geometric uniqueness} (or at least a version of it) of the initial value problem.

To show the above assertion is not an easy nor a direct task since Einstein's equations are geometrical and so they are subject to the diffeomorphism freedom. First, the field equations need to be recast into a system of evolution equations with constraints. Second, a gauge for which the first set of equations is strongly hyperbolic needs to be found. Third, the theory of strongly hyperbolic equations is used in order to show that a unique solution in this gauge class exists and depends continuously on the initial data. Finally, geometric uniqueness must be shown. This program was first accomplished using harmonic coordinates.\cite{cB51,cBrG69}

Contrary to the initial value problem the data at the boundary ${\cal T}$ is not unique. Here, we are thinking about the case where ${\cal T}$ represents a time-like surface (with respect to the unknown metric $g$). Therefore, what can be specified at the boundary is, essentially, the incoming radiation which could be given as an absolute quantity, zero for all times, say, or which could be given as a function of the outgoing radiation, as is the case for a mirror in electromagnetism, for instance. A further complication which makes the problem difficult to tackle arises from the absence of a reliable local definition for gravitational radiation in general relativity. Ultimately, this problem is related to the fact that the initial-boundary value problem (IBVP) we are considering appears as a means for having an artificial boundary for numerical calculations, and not an IBVP with physical boundaries which would, presumably, dictate the correct boundary conditions. 
This problem is, in fact, connected to two other ones: the first one is that in most formulations some of the constraint variables, that is, the quantities which yield the constraint equations when set to zero, have nonzero propagation speeds with respect to the boundary. In particular, there are constraints propagating into the computation domain, and in this case the boundary data must be chosen with care to ensure that no constraint-violating modes enter the domain. The second problem, which conspires with the first one in making the task difficult, is that in most formulations the problem is characteristic. We shall provide a precise definition shortly, but roughly speaking it means that there are perturbations having zero speeds with respect to the boundary, that is, modes which propagate along the boundary. It is hard to control such perturbations at the boundary, and so it is difficult to formulate a well posed problem if such zero speed perturbations appear in the boundary conditions fixing the constraints.

The IBVP for Einstein's vacuum field equations was solved by Friedrich and Nagy in 1999\cite{hFgN99} based on a frame formalism with very precise gauge conditions tailored to the problem. Furthermore, their formulation also modifies the evolution equations near the boundary in such a way that the constraint variables propagate along the boundary. This feature implies that no constraint conditions are needed at the boundary and one gets away with boundary conditions which only control the physically relevant field components. After this impressive work most attempts to solve the problem for metric based formulations failed, the reason being that several of the issues mentioned above still needed to be understood.

The first breakthrough came in 2006 when Kreiss and Winicour\cite{hKjW06} realized two very important points. The first point is that the second-order wave equations, when viewed as a certain pseudodifferential first-order operator, is not characteristic. This could be applied\cite{gNoOoR04} to Einstein's field equations in harmonic coordinates, in which case the evolution equations reduce to a quasilinear system of wave equations for the metric components. Second, they realized that it was possible to specify constraint-preserving boundary conditions  for this system similar to Sommerfeld boundary conditions, i.e. non-incoming radiation conditions, but with a certain coupling obeying a hierarchy which allowed to prove strong stability (defined below). More recently, the above results were also established using the usual energy estimates based on integration by parts.\cite{hKoRoSjW07} Furthermore, it was possible to conclude that the boundary conditions imposed were actually of the maximal dissipative type\cite{kF58,pLrP60} for a specific, non-standard class of first-order symmetric hyperbolic reductions for which the system is not characteristic.\cite{hKoRoSjW09} This also opens the interesting possibility to extend the result to discretizations of the equations using finite differences satisfying "summation by parts".\cite{GKO95}

The plan for the remainder of this paper is as follows. In the next section we discuss boundary conditions for strongly hyperbolic systems, introduce the concept of strongly stable problems and assert the main theorem on the subject together with two applications. In section~\ref{Sec:Uniqueness} we explore the boundary conditions for the linearized theory and analyze the geometric uniqueness problem. Finally, conclusions and open issues are discussed in section~\ref{Sec:Conclusions}.

\section{Strongly stable IBVP}

Consider a first-order quasilinear system of the form
\begin{equation}
\partial_t u^{\alpha} = A^{\alpha c}{}_{\beta}(u,x,t) \nabla_c u^{\beta} + B^{\alpha}(u,x,t),
\label{Eq:FirstOrderSystem}
\end{equation}
where $u^\alpha = u^\alpha(x,t)$, $\alpha=1,2,...N$, are the unknowns, $\nabla_c$ is a connection on $M$ and $A^{\alpha c}{}_{\beta}(u,x,t)$ and $B^{\alpha}(u,x,t)$ depend smoothly on their arguments. We say the above system is {\sl strongly hyperbolic} if there exists a {\sl symmetrizer}, that is, a symmetric, positive definite matrix $H_{\alpha\beta} = H(u,x,t,n)_{\alpha \beta}$, depending smoothly on its arguments, such that $H_{\alpha \gamma} A^{\gamma c}{}_{\beta} n_c$ is also symmetric for all $(u,x,t)$ and all one-forms $n_c$. The system is called {\sl symmetric hyperbolic} if, in addition, the symmetrizer can be chosen independent of $n$. A quasilinear first-order system has a well posed initial-value problem which is stable under zeroth order perturbations (changes in $B^{\alpha}(u,x,t)$) if and only if it is strongly hyperbolic. Most physical problems describing wave propagation, i.e. electromagnetism, fluids, elasticity, general relativity, etc, can be cast into first-order symmetric or strongly hyperbolic systems, see for instance Ref.~\cite{oR98} and references therein; in particular see Refs.~\cite{GerochPDEP} and \cite{oR04} for covariant definitions.

Modulo the smoothness assumptions strong hyperbolicity is equivalent to asking that for each $n_c$ the {\em principal symbol} $A^{\alpha c}{}_{\beta}(u,x,t) n_c$ has a complete set of eigenvectors with real eigenvalues. This implies that, locally, every high frequency solution is a combination of plane waves where the amplitude, frequency and wave vector of the wave is given by, respectively, the eigenvector, the eigenvalue and the one-form $n_c$ of the principal symbol. With the help of the symmetrizer it is possible to construct norms giving a priori energy estimates, a basic step in showing well posedness for the initial value problem.

When boundaries are present, a systematic method\footnote{See, for instance, Refs.~\cite{KL89} and \cite{GKO95} and references therein.} for constructing boundary conditions consists in analyzing the plane waves with normal incidence to the boundary, that is, in diagonalizing the {\sl boundary matrix}
\begin{displaymath}
N^\alpha{}_\beta(u,x,t) := A^{\alpha c}{}_{\beta}(u,x,t) n_c,
\end{displaymath}
where $n_c$ is an outgoing normal one-form to the boundary surface.\footnote{Notice that no metric is needed to determine the one-form $n_c$.} The field perturbations  $\delta u^\alpha$ can be written as a linear combination of the eigenvectors. The coefficients corresponding to positive, negative and zero eigenvalues are called the {\sl incoming}, {\sl outgoing} and {\sl zero speed} fields, respectively. One must specify a condition for each incoming field. In contrast to this, the outgoing fields are determined by the evolution of the solution up to the given time. The zero speed fields are also predetermined by the evolution, but unfortunately in a weaker sense in comparison with the outgoing ones, and this is the main problem of the theory. The presence of these zero speed fields, that is the existence of zero eigenvalues of the matrix $N$, is usually referred as a problem with {\sl characteristic boundary}. The boundary condition specifies data to the incoming fields, or to an appropriate linear combination of the in- and outgoing fields, as is the case for reflecting boundaries. However, the linear combination cannot include zero speed fields. Unfortunately, in theories with constraints, the boundary conditions required to enforce the constraints do usually contain such zero speed fields.

\subsection{Example: Maxwell's equations in the potential version}
\label{SubSec:MaxwellFirstOrder}

In order to illustrate this problem we consider Maxwell's equations in the Lorentz gauge, in which case on obtains a set of wave equations for the components of the vector potential $A_b$. These equations are only coupled among them by the constraint the gauge imposes, namely the requirement for $A_b$ to be divergence-free. For simplicity we just consider the theory in Minkowski spacetime, in which case
\begin{displaymath}
\nabla^a\nabla_a A_b = 0,\qquad
\nabla^b A_b = 0.
\end{displaymath}
Introducing the first-order derivatives of $A_b$ as new independent fields, $D_{ab} := \nabla_a A_b$, the set of wave equations is reducible to the first-order system
\begin{displaymath}
\partial_t A_b = D_{tb},\qquad
\partial_t D_{tb} = \partial^j D_{jb},\qquad
\partial_t D_{jb} = \partial_j D_{tb},
\end{displaymath}
where $j=x,y,z$. Denoting by $u^\alpha$ the $20$-component vector $(A_b,D_{tb},D_{jb})$, this system has the form of Eq.~(\ref{Eq:FirstOrderSystem}) where the principal symbol
\begin{displaymath}
A^{\alpha i}{}_\beta n_i u^\beta = (0,n^j D_{jb}, n_j D_{tb})
\end{displaymath}
is symmetrized by the trivial symmetrizer $H_{\alpha\beta} = \delta_{\alpha\beta}$. Therefore, the Cauchy problem for this system is well posed. If initial data is given for $A_b$ and its time derivative, both being divergence-free, then a unique solution exists, and it depends continuously on the given data.

Next, let us consider the system on the half space $\Sigma := \{ (x,y,z)\in\Real^3 : x > 0 \}$. The normal matrix is obtained from the principal symbol with $n = -dx$ the unit normal one-form. Therefore, the zero, in- and outgoing fields are
\begin{displaymath}
\begin{array}{rl}
D_{tb} + D_{xb} = (\partial_t + \partial_x)A_b &  \mbox{(incoming field)}\\
D_{tb} - D_{xb} = (\partial_t - \partial_x)A_b &  \mbox{(outgoing field)}\\
D_{yb} = \partial_y A_b  & \mbox{(zero speed field)}\\
D_{zb} = \partial_z A_b  & \mbox{(zero speed field)}
\end{array}
\qquad
\partial_t A_t = \partial_x A_x + \partial_y A_y + \partial_z A_z,
\end{displaymath}
where on the right-hand side we have explicitly written down the constraint equation in order to explain the problem: namely, when written in terms of the eigenfields, we see that the constraint not only depends on the in- and outgoing fields, but also on the zero speed fields! Therefore, imposing the constraint on the boundary in order to guarantee constraint preservation leads to a boundary condition which couples the incoming fields to outgoing and zero speed fields.\footnote{Instead of imposing the constraint itself on the boundary one might try to set some linear combination of its normal and time derivatives to zero, obtaining a constraint-preserving boundary condition that does not involve zero speed fields. Unfortunately, this trick only seems to work for reflecting boundaries, see Refs.~\cite{bSjW03} and \cite{gCjPoRoSmT03} for the case of general relativity. In our example, such boundary conditions are given by $\partial_t A_t = \partial_x A_x = \partial_t A_y = \partial_t A_z = 0$ which imply $\partial_t(\nabla^b A_b) = 0$.} 

Of course, this problem does not arise when writing Maxwell's equations in terms of electric and magnetic fields, which directly leads to a symmetric hyperbolic system with trivial constraint propagation. So the fact that the system is characteristic and has constraints with nontrivial propagation has conspired to convert a simple problem into an intractable one. Notice also that in the potential formulation there is no clear correspondence between the real radiation degrees of freedom and the in- and outgoing fields. Some of the fields describe gauge degrees of freedom.

Until recently, the above problem was open.

\subsection{Well posed and strongly well-posed boundary conditions}

If one writes the equations as a first-order system, then at an initial surface the value of all the fields needs to be prescribed (in such a way that they satisfy the constraints if such are present) in order to obtain a unique solution. At the boundary the situation is different in general. Only data for the incoming fields must be given while the values the other fields acquire depend on the evolution in the bulk. Then, the question is whether or not one can control the fields in terms of the given initial and boundary data when imposing a boundary condition which couples the incoming fields with the outgoing and zero speed fields. Here, by "control" we mean the following:

\begin{definition}
\label{Def:StrongStab}
Consider a first-order strongly hyperbolic system of the form (\ref{Eq:FirstOrderSystem}) with initial data $\left. u^\alpha \right|_{\Sigma_0} = f^\alpha$ and boundary conditions of the form $\left. L^{\Gamma}{}_\alpha(t,x,u) u^{\alpha} \right|_{\cal T} = g^{\Gamma}$. This problem is called {\em strongly stable}\cite{KL89} if there are norms $\|\cdot \|_{\Sigma}$, $\|\cdot\|_{\cal T}$ bounding the fields and their derivatives on $\Sigma_T$ and ${\cal T}$, respectively, and a constant $\eps > 0$ such that each solution $u$ satisfies an estimate of the form
\begin{equation}
\| u \|_{\Sigma_{T}} + {\eps} \|u \|_{\cal{T}} \leq F(T,\| f \|_{\Sigma_0}, \| g \|_{\cal{T}}),
\label{Eq:WPEstimate}
\end{equation}
where $F(t,x,y)$ is a smooth function in its arguments that vanishes when $t=0$ or when $(x,y)=(0,0)$.
\end{definition}

\begin{remark}
In particular, the estimate~(\ref{Eq:WPEstimate}) implies the continuous dependence of the solutions on their data. A problem which is strongly stable and for which a unique solution exists on a suitable time interval $[0,T]$ is called {\em strongly well posed}.
\end{remark}

A trivial example of a strongly stable problem is a symmetric hyperbolic system for which all the fields are incoming at the boundary, since in this case the values for all the fields must be specified there. More generally, a symmetric hyperbolic system with non-characteristic boundary surface and maximal dissipative boundary conditions\cite{kF58,pLrP60} is strongly stable. A general theory for strictly hyperbolic first-order systems which allows to determine which boundary conditions lead to a strongly stable problem by purely algebraic methods was developed by Kreiss\cite{hK70}.

If a problem satisfies the same properties as in Def.~\ref{Def:StrongStab} with the difference that the constant $\eps$ in (\ref{Eq:WPEstimate}) is zero, then we say that the problem is {\em stable}. In this case, the estimate (\ref{Eq:WPEstimate}) can still be used to show existence and uniqueness properties for some systems, but the set of systems
for which this can be done is rather small, usually involving only linear systems or very restrictive boundary conditions. For this reason, in the following, we restrict ourselves to a smaller class of strongly hyperbolic systems for which strong well posedness can be shown.


\subsection{Second order systems}

A particularly interesting set of systems are those which are second-order, in particular those who describe systems of wave equations of the form
\begin{equation}
g^{ab}(\Phi^B)\nabla_a\nabla_b \Phi^A + F^A(\Phi^B, \nabla_c \Phi^C) = 0,\qquad
A=1...N,
\label{Eq:NonLinearWave}
\end{equation} 
where $\Phi^A$ is some set of tensorial fields, $g^{ab}$ a Lorentzian metric which in general depends locally on the fields, and $\nabla$ is a given connection on $M$. We impose initial conditions,
\begin{equation}
\Phi^A|_{\Sigma_0} = \Phi_0^A, \qquad 
\left. \partial_t\Phi^A \right|_{\Sigma_0} = \Pi_0^A
\label{Eq:WaveID}
\end{equation}
and boundary condition of the following form:
\begin{equation}
(T^d + a N^d)\nabla_d\Phi^A
 = \sum_{B=1}^N c^{Ad}{}_B(t,x,\Phi)\nabla_d \Phi^B 
 + \sum_{B=1}^N d^A{}_B(t,x,\Phi)\Phi^B + G^A(t,x),
\label{Eq:WaveBC}
\end{equation}
where here $a > 0$, $T$ is a future-directed time-like vector field tangent to ${\cal T}$ and $N$ is the outward unit normal to ${\cal T}$, see Fig.~\ref{Fig:IBVP}. Furthermore, the matrix coefficients $c^{Ad}{}_B$ have the property that they can be made arbitrarily small by an appropriate local linear transformation $\Phi^A \to J^A{}_B\Phi^B$ of the fields (see Ref.~\cite{hKoRoSjW09} for the details). In particular, this is the case if the matrix operator $c^{Ad}{}_B\nabla_d$ is in upper triangular form with zeroes on the diagonal since then the matrix elements $c^{Ad}{}_B$ can be made arbitrarily small by a simple rescaling of the fields, much like in the proof of the Lyapunov stability theorem. We have the following theorem:\cite{hKjW06,hKoRoSjW07,hKoRoSjW09}

\begin{theorem}
\label{Thm:Main}
The IBVP (\ref{Eq:NonLinearWave},\ref{Eq:WaveID},\ref{Eq:WaveBC}) is strongly well posed.
\end{theorem}

This theorem constitutes the key result which allows the consideration of a whole new class of boundary conditions which are flexible enough to solve the IBVP of electromagnetism in the potential formulation and also the full Einstein vacuum field equations in harmonic coordinates.\cite{hKjW06,hKoRoSjW09} This new result was obtained by realizing two things: first, that the above equations, when viewed as a pseudodifferential system\footnote{Think of this as the Fourier transform of the system.} and then reduced to first-order is a non-characteristic system (see for instance Refs.~\cite{KreissOrtiz-2002} and \cite{gNoOoR04}), and second, that the Kreiss theory\cite{hK70} applies equally well to pseudodifferential equations. The above boundary conditions are just of the type required in this theory to give a strongly well posed system. Later, the problem has been understood from a different point of view using more mundane energy estimates based on integration by parts instead of pseudodifferential calculus.\cite{hKoRoSjW07} The key observation is that for second-order systems like Eq.~(\ref{Eq:NonLinearWave}) there are many different hyperbolizations, and so many different energy norms. This freedom comes about because for a general Lorentzian metric there is no preferred time direction. Different time directions give rise to first-order reductions which differ among each others by constraint terms, and so they are inequivalent. By choosing the time direction in a suitable, but nonstandard form one obtains a non-characteristic first-order system which is amenable with the usual theory. 
Notice that this situation is rather different than the one in fluid dynamics, where the four-velocity singles out a preferred time direction at each point of spacetime.


\subsection{Application: Maxwell's equations in the potential version}
\label{SubSec:MaxwellSecondOrder}

As a first application, we return to Maxwell's equations for the four-vector potential on the half space $\Sigma$,
\begin{displaymath}
\nabla^a\nabla_a A_b = 0,\qquad
\nabla^b A_b = 0.
\end{displaymath}
The initial data $\left. A_b \right|_{\Sigma_0} = f_b^{(0)}$, $\left. \partial_t A_b \right|_{\Sigma_0} = f_b^{(1)}$ is subject to the conditions $\nabla^b f_b^{(0)} = 0$ and $\nabla^b f_b^{(1)} = 0$, so that the constraints are correctly propagated into the domain of dependence of the initial surface $\Sigma_0$. The boundary conditions are described in  a geometrically elegant way by introducing the complex null basis of vector fields
\begin{displaymath}
K := \partial_t - \partial_x,\quad
L:=\partial_t + \partial_x,\quad
Q:=\partial_y + i\partial_z,\quad
\bar{Q}:=\partial_y - i\partial_z,\qquad
i = \sqrt{-1}.
\end{displaymath}
Then, we consider the following set of conditions at the boundary ${\cal{T}} = [0,T]\times\partial\Sigma$,
\begin{equation}
\begin{array}{ll}
\nabla_K A_K = q_K, & \hbox{(gauge)}\\
\nabla_K A_Q = q'_Q & \hbox{(Sommerfeld)}\\
\nabla_K A_L = -\nabla_L A_K + \nabla_Q A_{\bar{Q}} + \nabla_{\bar{Q}} A_Q, &
\hbox{(constraint $\nabla^b A_b = 0$)}
\end{array}
\label{Eq:MaxwellBC1}
\end{equation}
where $A_K := K^b A_b$, $\nabla_K := K^a\nabla_a$, etc... and $q_K$, $q'_Q$ are boundary data. The first two equations (which constitute three real conditions since $Q$ is complex) are non-incoming conditions, usually called Sommerfeld conditions, since they control the incoming fields. The left-hand side of the third equation is also an incoming field, but now it is sourced by first-order derivatives of the fields $A_K$ and $A_Q$, for which the first two equations already give conditions to. Therefore, this system conforms with the hypothesis of Theorem~\ref{Thm:Main} with $a=1$, $T=\partial_t$ and $N=-\partial_x$ such that $T + a N = K$ and leads to a strongly well posed problem. A simple way of understanding why this works for the second-order problem whereas it failed for the first-order reduction discussed in subsection~\ref{SubSec:MaxwellFirstOrder} is the following: the first two equations in the boundary conditions~(\ref{Eq:MaxwellBC1}) give rise to a strongly stable system for $A_K$ and $A_Q$, respectively, and hence $A_K$ and $A_Q$ each satisfy an estimate of the form (\ref{Eq:WPEstimate}). 
In particular, one controls all first derivatives of the fields at the boundary. Therefore, when applying the estimate to the wave equation for $A_L$ with the third boundary conditions, the source terms on the right-hand side are controlled.

Regarding the physical interpretation of the first two conditions in Eq.~(\ref{Eq:MaxwellBC1}), the first controls the incoming gauge fields of the form $A_b = \nabla_b(e^{i L\cdot x})$ while the second controls incoming electromagnetic fields of the form $A_b = \bar{Q}_b e^{i L\cdot x}$. However, a problem arises because $\nabla_K A_Q$ is not gauge-invariant. For this reason, we replace the boundary conditions~(\ref{Eq:MaxwellBC1}) by
\begin{equation}
\begin{array}{ll}
\nabla_K A_K = q_K, & \hbox{(gauge)}\\
\nabla_K A_Q - \nabla_Q A_K = q_Q & \hbox{(radiation)}\\
\nabla_K A_L = -\nabla_L A_K + \nabla_Q A_{\bar{Q}} + \nabla_{\bar{Q}} A_Q, &
\hbox{(constraint $\nabla^b A_b = 0$)}
\end{array}
\label{Eq:MaxwellBC2}
\end{equation}
where now the boundary data $q_Q$ represents the $KQ$-components of the electromagnetic field tensor which is gauge-invariant. 
The new boundary conditions~(\ref{Eq:MaxwellBC2}) still satisfy the hypothesis of Theorem~\ref{Thm:Main} and we obtain a strongly well posed problem. 
As we shall see, these new boundary conditions lead to gauge uniqueness, whereas the conditions~(\ref{Eq:MaxwellBC1}) do not.

\subsection{Application: The linearized Einstein equations in harmonic coordinates}
\label{SubSec:EinsteinSecondOrder}

As a second application we describe a strongly well posed IBVP for the Einstein field equations in vacuum when linearized on the flat background $(M,\eta)$ where $M = [0,\infty) \times \Sigma$, $\Sigma$ denoting half space and $\eta = -dt^2 + dx^2 + dy^2 + dz^2$ the Minkowksi metric. The restrictions of linearization and working on half space are done for the sake of simplicity. A (local in time) well posed formulation for the full vacuum equations is given in Ref.~\cite{hKoRoSjW09}. When harmonic coordinates are used the linearized Einstein equations are
\begin{displaymath}
\nabla^c\nabla_c h_{ab} = 0,\qquad
H_a := \nabla^b\left( h_{ab} - \frac{1}{2}\eta_{ab}\eta^{cd} h_{cd} \right) = 0,
\end{displaymath}
where $h_{ab}$ denotes the first variation of the metric fields. As in the previous application, the initial data $\left. h_{ab} \right|_{\Sigma_0} = f_{ab}^{(0)}$, $\left. \partial_t h_{ab} \right|_{\Sigma_0} = f_{ab}^{(1)}$ is chosen such that the constraints $H_a = 0$ and $\partial_t H_a = 0$ are satisfied at $\Sigma_0$. In order to describe the boundary conditions, we use the complex null basis of vector fields $K$, $L$, $Q$, $\bar{Q}$ defined in the previous application. Let us start with the constraint conditions and work our way up on the triangular structure of the boundary conditions. The  imposition of the constraints $H_a = 0$ at the boundary surface ${\cal T}$ yields
\begin{eqnarray}
\nabla_K h_{Q\bar{Q}} 
 &=& -\nabla_L h_{KK} + \nabla_Q h_{K\bar{Q}} + \nabla_{\bar{Q}} h_{KQ},
\nonumber\\
\nabla_K h_{LQ} 
 &=& -\nabla_L h_{KQ} + \nabla_Q h_{KL} + \nabla_{\bar{Q}} h_{QQ},
\label{Eq:EinsteinCPBC}\\
 \nabla_K h_{LL} 
 &=& -\nabla_L h_{Q\bar{Q}} + \nabla_Q h_{L\bar{Q}} + \nabla_{\bar{Q}} h_{LQ}.
\nonumber
\end{eqnarray}
The right-hand side of the last equation contains first-order derivatives of the fields $h_{Q\bar{Q}}$ and $h_{LQ}$ which are controlled in the first two equations. The right-hand sides of the first two equations require control of the first-order derivatives of $h_{KK}$, $h_{KQ}$, $h_{KL}$ and $h_{QQ}$. Therefore, we impose the following gauge boundary conditions:
\begin{eqnarray}
\nabla_K h_{KK} &=& q_{K},
\nonumber\\
\nabla_K h_{KQ} - \frac{1}{2}\nabla_Q h_{KK}  &=& q_{Q},
\label{Eq:EinsteinGaugeBC}\\
\nabla_K h_{KL} - \frac{1}{2}\nabla_L h_{KK} &=& q_{L},
\nonumber
\end{eqnarray}
which control the fields $h_{KK}$, $h_{KQ}$ and $h_{KL}$. Here, $q_K$, $q_Q$ and $q_L$ are boundary data, and the purpose of introducing the second term on the left-hand sides of the last two equations will become clear in the next section. It remains to control the field $h_{QQ}$, which is related to the field $h_{ab} = \bar{Q}_a\bar{Q}_b e^{i L\cdot x}$ describing incoming gravitational radiation. Two possibilities have been contemplated for controlling this field. The first specifies the shear, $\sigma := Q^a Q^b\nabla_a K_b$, of the null congruence corresponding to the outgoing null vector field $K$.\cite{mRoRoS07,jW09a,jW09b} In the linearized setting considered here this yields the condition
\begin{equation}
\nabla_K h_{QQ} - 2\nabla_Q h_{KQ} = 2\sigma.
\label{Eq:EinsteinShearBC}
\end{equation}
The boundary conditions~(\ref{Eq:EinsteinGaugeBC},\ref{Eq:EinsteinShearBC},\ref{Eq:EinsteinCPBC}) satisfy the hypothesis of Theorem~\ref{Thm:Main}, and therefore, one obtains a strongly well posed IBVP.

However, in the next section we will see that the shear condition~(\ref{Eq:EinsteinShearBC}) leads to difficulties when discussing geometric uniqueness. For this reason, it is convenient to replace this condition by a new condition which specifies the Weyl scalar $\psi_0 := -2 K^a Q^b K^c Q^d R_{abcd}$ instead, where here, $R_{abcd}$ denotes the linearized curvature tensor associated to $h_{ab}$. This gives\cite{lLmSlKrOoR06,mRoRoS07}
\begin{equation}
\nabla_K^2 h_{QQ} + \nabla_Q(\nabla_Q h_{KK} - 2\nabla_K h_{KQ}) = \psi_0.
\label{Eq:EinsteinWeylBC}
\end{equation}
This condition involves second-order derivatives of the fields and handling it requires a little bit more work. However, using the pseudodifferential first-order reduction of Ref.~\cite{hKjW06} it is possible to prove that the boundary conditions~(\ref{Eq:EinsteinGaugeBC},\ref{Eq:EinsteinWeylBC},\ref{Eq:EinsteinCPBC}) lead to a strongly well posed IBVP, see Ref.~\cite{mRoRoS07}. Estimates on the amount of spurious reflections introduced by the boundary conditions~(\ref{Eq:EinsteinShearBC}) and (\ref{Eq:EinsteinWeylBC}) were given in Refs.~\cite{lBoS06}, \cite{oRlLmS07} and \cite{mRoRoS07}.

\section{Geometric uniqueness}
\label{Sec:Uniqueness}

In the previous section we have reviewed well posed IBVP for Maxwell's equations and the linearized Einstein's equations. These problems allow, from the point of view of partial differential equations, to construct unique solutions given appropriate initial and boundary data. In this section we discuss the question of gauge uniqueness, i.e. the relation that initial and boundary data need to satisfy such that the solutions they give rise to are gauge related.

As discussed in the introduction, two initial data sets $(h,k)$ and $(\tilde{h},\tilde{k})$ on $\Sigma_0$ are related to each other by a diffeomorphism of $\Sigma_0$ if and only if
their corresponding Cauchy developments on $D(\Sigma_0)$ are related to each other by a diffeomorphism of $D(\Sigma_0)$ which leaves $\Sigma_0$ invariant. It would be nice to have a similar statement for the IBVP, that is a similar statement that does not only consider two solutions on the domain of dependence $D(\Sigma_0)$ of the initial surface but on the whole manifold $M = [0,T]\times \Sigma$. Given two initial data sets $(h,k)$ and $(\tilde{h},\tilde{k})$ on $\Sigma_0$ and two boundary data sets $q$ and $\tilde{q}$ on ${\cal T} := [0,T]\times\partial\Sigma$ satisfying suitable compatibility conditions at the edge $S:= \{0 \} \times \partial\Sigma$, one would like to know under which circumstances the corresponding solutions $(M,g)$ and $(M,\tilde{g})$ are related to each other by a diffeomorphism on $M$ which leaves $\Sigma_0$ and ${\cal T}$ invariant. Is it, say,  possible to relate $q$ and $\tilde{q}$ by a transformation on ${\cal T}$ alone, such that for given initial data on $\Sigma_0$ the resulting metrics $g$ and $\tilde{g}$ are related to each other by a diffeomorphism?

As pointed out in Ref.~\cite{hF09} there are several difficulties with this question:
\begin{enumerate}
\item[(i)] It is a priori not clear what the boundary data $q$ should be. Unlike for the case of the initial surface, $q$ cannot represent the first and second fundamental forms of ${\cal T}$, as explained in the introduction.
\item[(ii)] The boundary data $q = (q_K,q_Q,q_L,\sigma)$ and $q = (q_K,q_Q,q_L,\psi)$, respectively, introduced in subsection~\ref{SubSec:EinsteinSecondOrder} depend on a specific choice of the outgoing null vector field $K$, or at least on its orthonormal projection onto the boundary surface which is a time-like vector field $T$. It is not clear if in general there is such a preferred time direction on ${\cal T}$.\footnote{However, see the recent proposal in Ref.~\cite{hF09}.}
\item[(iii)] Related to the first two items is the question whether or not it is at all possible to find a transformation on ${\cal T}$ which relates $q$ and $\tilde{q}$, independent of the initial data.
\end{enumerate}
One possibility for dealing with these issues is to introduce a background metric and to formulate the boundary conditions in a covariant way based on the covariant derivative defined by the background metric. This is the approach taken in Refs.~\cite{mRoRoS07} and \cite{hKoRoSjW09} and further developed in Refs.~\cite{jW09a} and \cite{jW09b}. However, the issues above are then traded by the question on the dependency of the solution on the background metric.

In the following, we analyze these issues in the much simpler setting of Maxwell's equations in the potential formulation, where the role of the diffeomorphism is played by the gauge transformations, and linearized gravity.

\subsection{Gauge uniqueness for Maxwell's equations for the potential formulation}

We reconsider the source free Maxwell equation for the four-vector potential $A_b$ on the half-space $\Sigma$,
\begin{equation}
\nabla^a(\nabla_a A_b - \nabla_b A_a) = 0.
\label{Eq:Maxwell}
\end{equation}
In contrast to the previous examples, however, we do not necessarily impose the Lorentz gauge. The initial data for this problem are the three-vector potential and the electric field, $(A^{(0)}_j,E^{(0)}_j)$, and the boundary data $q_Q$ which corresponds to the $KQ$-components of the electromagnetic field tensor and is related to the incoming electromagnetic radiation in the direction of $L = \partial_t + \partial_x$, see subsection~\ref{SubSec:MaxwellSecondOrder}. We assume that this data is smooth, satisfies the Gauss constraint $\partial^j E^{(0)}_j = 0$ and suitable compatibility conditions at $S$. Since the initial value of the magnetic field only determines $A^{(0)}_j$ up to the addition of a gradient, we are looking for a solution of Eq.~(\ref{Eq:Maxwell}) such that
\begin{equation}
\left. A_j \right|_{\Sigma_0} = A^{(0)}_j + \partial_j\chi^{(0)},\qquad
\left. \partial_t A_j - \partial_j A_0 \right|_{\Sigma_0} = E^{(0)}_j,\qquad
\left. \nabla_K A_Q - \nabla_Q A_K \right|_{\cal T} = q_Q,
\label{Eq:MaxwellData}
\end{equation}
for some smooth function $\chi^{(0)}$ on $\Sigma_0$. The next result shows that this data determines a smooth solution $A_b$ of Maxwell's equations which is unique up to a gauge 
transformation.

\begin{theorem}
The IBVP (\ref{Eq:Maxwell},\ref{Eq:MaxwellData}) possesses a smooth solution $A_b$ which is unique up to a gauge transformation $\tilde{A}_b = A_b + \nabla_b\chi$.
\end{theorem}

{\bf Proof}: We first show the existence of a solution in the Lorentz gauge, for which Eq.~(\ref{Eq:Maxwell}) reduces to the system of wave equations $\nabla^a\nabla_a A_b = 0$. 
The initial data, $(A_b|_{\Sigma_0},\partial_t A_b|_{\Sigma_0})$, is chosen such that 
$\left. A_j \right|_{\Sigma_0} = A^{(0)}_j$, $\left. \partial_t A_j \right|_{\Sigma_0} = E_j^{(0)} + \left. \partial_j A_0 \right|_{\Sigma_0}$ and $\left. \partial_t A_0 \right|_{\Sigma_0} = \partial^j A_j^{(0)}$, where the initial data for $A_0$ is chosen smooth but otherwise arbitrarily. Then, we evolve the well-posed IBVP with boundary conditions~(\ref{Eq:MaxwellBC2}) described in subsection~\ref{SubSec:MaxwellSecondOrder}, where the source function $q_K$ is chosen to be smooth and to satisfy the compatibility conditions at $S$, but otherwise is arbitrary. By construction, the resulting solution $A_b$ solves $\left. \nabla^b A_b \right|_{\Sigma_0} = 0$, $\left. \partial_t\nabla^b A_b \right|_{\Sigma_0} = 0$, and therefore, it yields a solution of Maxwell's equations~(\ref{Eq:Maxwell}) in the Lorentz gauge with data (\ref{Eq:MaxwellData}) for $\chi^{(0)}=0$.

As for uniqueness, let $A_b^{(1)}$ and $A_b^{(2)}$ be two smooth solutions of Eqs.~(\ref{Eq:Maxwell},\ref{Eq:MaxwellData}). We show there exists a function $\chi$ on $M$ such that $\delta A_b := A_b^{(2)} - A_b^{(1)} = \nabla_b\chi$, that is, the two solutions are gauge related on $M$. In order to see this, we first notice that $\delta A_b$ satisfies Maxwell's equations~(\ref{Eq:Maxwell}) with initial data satisfying $\left. \delta A_j \right|_{\Sigma_0} = \partial_j\delta\chi^{(0)}$ and $\left. \partial_t \delta A_j - \partial_t \delta A_0 \right|_{\Sigma_0} = 0$ and trivial boundary data, $q_Q = 0$. Next, we transform $\delta A_b$ into the Lorentz gauge by finding $\chi$ such that $\nabla^b\nabla_b\chi = \nabla^b \delta A_b$ and setting $\tilde{A}_b := \delta A_b - \nabla_b\chi$. The transformed potential $\tilde{A}_b$ satisfies the Lorentz gauge. Furthermore, choosing initial and boundary data for $\chi$ such that $\left. \chi\right|_{\Sigma_0} = \delta\chi^{(0)}$,
 $\left. \partial_t\chi\right|_{\Sigma_0} = \left. \delta A_0 \right|_{\Sigma_0}$, 
 and $\left. \nabla_K\nabla_K\chi \right|_{\cal T} = \left. \nabla_K A_K \right|_{\cal T}$ we obtain $\left.\tilde{A}_b \right|_{\Sigma_0} = 0$, $\left. \partial_t\tilde{A}_b \right|_{\Sigma_0} = 0$ and $\left. \nabla_K\tilde{A}_K \right|_{\cal T} = 0$. Therefore, $\tilde{A}_b$ is a smooth solution of the IBVP described in subsection~\ref{SubSec:MaxwellSecondOrder} which has trivial initial data and satisfies the boundary conditions~(\ref{Eq:MaxwellBC2}) with trivial data. By uniqueness of this problem it follows that $\tilde{A}_b = 0$, which implies that $\delta A_b = \nabla_b\chi$.
\qed

\begin{remark}
From the proof of the theorem it follows that the choice for $\left. A_0\right|_{\Sigma_0}$ and the boundary data $q_K$ in the IBVP described in subsection~\ref{SubSec:MaxwellSecondOrder} with boundary conditions~(\ref{Eq:MaxwellBC2})  corresponds to a pure gauge freedom.

In contrast to this, the same IBVP with boundary conditions~(\ref{Eq:MaxwellBC2}) replaced by the gauge-dependent condition~(\ref{Eq:MaxwellBC1}), $\left. \nabla_K A_Q \right|_{\cal T} = q'_Q$, does not lead to gauge uniqueness. The reason is that a gauge transformation induces the transformations $\tilde{q}_K = q_K + \nabla_K\nabla_K\chi$ and $\tilde{q}'_Q = q'_Q + \nabla_K\nabla_Q\chi$ on the boundary data, which overdetermines the gauge function $\chi$.
\end{remark}

\subsection{Geometric uniqueness for the linearized Einstein equations}

Einstein's field equations in vacuum, when linearized about the Minkowski metric $\eta_{ab}$, are
\begin{equation}
-\nabla^c\nabla_c h_{ab} - \nabla_a\nabla_b(\eta^{cd} h_{cd}) 
 + 2\nabla^c\nabla_{(a} h_{b)c} = 0,
\label{Eq:LinEinstein}
\end{equation}
where $h_{ab}$ denotes the first variation of the metric. The initial and boundary data are 
the first and second fundamental form of the initial surface, $(h^{(0)}_{ij},k^{(0)}_{ij})$, and the Weyl scalar $\psi_0$ defined in Eq.~(\ref{Eq:EinsteinWeylBC}), respectively. We assume the data to be smooth, to satisfy suitable compatibility conditions at $S$, and to satisfy the linearized Hamiltonian and momentum constraints $G^{ijrs}\partial_i\partial_j h^{(0)}_{rs} = 0$ and $G^{ijrs}\partial_j k^{(0)}_{rs} = 0$, where $G^{ijrs} := \delta^{i(r}\delta^{s)j} - \delta^{ij}\delta^{rs}$. We are looking for a solution of Eq.~(\ref{Eq:LinEinstein}) satisfying
\begin{eqnarray}
&& \left. h_{ij} \right|_{\Sigma_0} = h^{(0)}_{ij} + 2\partial_{(i} X_{j)},\qquad
\left. \partial_t h_{ij} - 2\partial_{(i} h_{j)0} \right|_{\Sigma_0} 
 = -2(k^{(0)}_{ij} + \partial_i \partial_j f),
\nonumber\\
&& \left. \nabla_K^2 h_{QQ} + \nabla_Q(\nabla_Q h_{KK} - 2\nabla_K h_{KQ}) 
 \right|_{\cal T} = \psi_0,
\label{Eq:LinEinsteinData}
\end{eqnarray}
where $X_j$ and $f$ are a smooth vector field and a smooth function on $\Sigma_0$, respectively, representing the initial gauge freedom.

\begin{theorem}
The IBVP (\ref{Eq:LinEinstein},\ref{Eq:LinEinsteinData}) possesses a smooth solution $h_{ab}$ which is unique up to an infinitesimal coordinate transformation $\tilde{h}_{ab} = h_{ab} + 2\nabla_{(a}\xi_{b)}$ generated by a vector field $\xi_a$.
\end{theorem}

{\bf Proof}: The proof works exactly as in the electromagnetic case, and is based on the harmonic IBVP described in subsection~\ref{SubSec:EinsteinSecondOrder} with boundary conditions~(\ref{Eq:EinsteinWeylBC}).  
\qed

\begin{remark}
With respect to an infinitesimal coordinate transformation the boundary data $q_K$, $q_Q$ and $q_L$ introduced in subsection~\ref{SubSec:EinsteinSecondOrder} transforms according to
\begin{displaymath}
\tilde{q}_K - q_K = 2\nabla_K^2\xi_K,\qquad
\tilde{q}_Q - q_Q = \nabla_K^2\xi_Q,\qquad
\tilde{q}_L - q_L = \nabla_K^2\xi_L.
\end{displaymath}
The right-hand sides of these equations provide a complete set of boundary data for the gauge source vector $\xi_a$. Therefore, the choice for the data $\left. h_{00} \right|_{\Sigma_0}$, $\left. h_{0j} \right|_{\Sigma_0}$ and $q_K$, $q_Q$ and $q_L$ which is left unspecified by the physical data $(h^{(0)}_{ij},k^{(0)}_{ij})$ and $\psi_0$ in the harmonic IBVP corresponds to infinitesimal coordinate transformations.

On the other hand, the shear $\sigma$ transforms as $\tilde{\sigma} = \sigma - \nabla_Q^2\xi_K$. Therefore, replacing the boundary condition~(\ref{Eq:EinsteinWeylBC}) with the shear boundary condition~(\ref{Eq:EinsteinShearBC}) overdetermines $\xi_a$ at the boundary, and one does not obtain geometric uniqueness.
\end{remark}

Returning to the issues (i)--(iii) described above, we can say that in the simple case of linearization about Minkowski space we have solved the points (i) and (iii). That is, we have identified the boundary data for this problem that leads to geometric uniqueness. Regarding point (ii), the time-like vectors $T$ that we have chosen in our problem correspond to the future-directed unit normal to the initial surface $\Sigma_0$ at $S$, parallel transported along the geodesics orthogonal to $\Sigma_0$. It is clear that when considering the nonlinear case, or even for the case of linearization about a nonflat spacetime, several difficulties appear. These additional difficulties are mentioned in the next section.

\section{Conclusions and open issues}
\label{Sec:Conclusions}

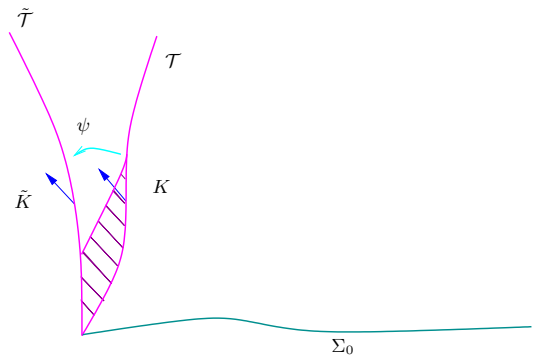
\begin{figure}[ht]
\begin{center}
    \resizebox{7cm}{!}{\input{zip_4.pstex_t}}
    \caption{Illustration of the zipper problem.}
    \label{Fig:Zipper}
\end{center}
\end{figure}

In spite of our better understanding of the IBVP for Einstein's field equations from the point of view of partial differential equations, several problems are still open and they have mostly a geometrical character. Geometric uniqueness presents at least two main obstacles. One is the question of how to generalize the local-linearized version we have discussed above to the global, fully nonlinear case. The problem is that we have imposed no restrictions on the normal component $\xi_N$ of the vector field generating the infinitesimal coordinate transformation. However, such a restriction is necessary in order to keep the boundary surface fixed under a diffeomorphism. Unfortunately, it does not seem possible to restrict $\xi_N$ with our current boundary conditions. As a consequence, the evolution of the boundary (viewed as an embedded surface in the unknown spacetime $(M,g)$) might depend on the initial gauge, say. Actually, a similar question arises when two solutions are considered on two different Cauchy surfaces and one asks the question whether or not the two solutions are related by a diffeomorphism. In order to answer this question, one needs to find the corresponding Cauchy developments. Our problem has this sort of difficulty also, and one could imagine solving this aspect of the problem "along the way", that is while one is computing the time evolution, one can ask, step by step, whether or not a diffeomorphism relating the two solutions exist. This is what we call the zipper problem. One is mending the gap between the two boundaries as one is solving the evolution equations, see Fig.~\ref{Fig:Zipper}.

The second obstacle is the choice of a time-like vector field $T$ in our present boundary conditions. Contrary to our initial expectation this problem might not be of a fundamental character; it is rather a manifestation of our inability to specify a non-incoming radiation condition correctly. Indeed, our present conditions single out the outgoing waves with wave vector along the null direction $K$ we have chosen. In this sense, our boundary condition is perfectly absorbing for such outgoing waves, but not for outgoing waves with different wave vectors. A genuine non-incoming wave boundary condition should be independent of any specific null or time-like direction at the boundary, and can only depend on the normal vector to it. For simpler systems like the scalar wave equation this is indeed the case:\cite{bEaM77} the imposition of perfectly absorbing boundary conditions leads to a nonlocal condition which is independent of a preferred time direction at the boundary. Although this condition might not be practical because of its nonlocality, it is possible to derive a hierarchy of local boundary conditions depending on a null vector $K$ approximating the nonlocal condition with better and better accuracy. So it is expected that the choice of $K$ becomes less and less relevant as one moves down the hierarchy. It should be interesting to explore this idea for the case of general relativity, departing from a nonlocal boundary condition which is independent of $K$.\footnote{See Ref.~\cite{oS07} for a recent review on absorbing local boundary conditions in general relativity.}

We end this article by asking whether or not we can apply the methods discussed here to metric formulations of Einstein's field equations other than the harmonic one. For example, a formulation that is widely used in numerical calculations is the BSSN formulation\cite{mStN95,tBsS98} which yields a mixed first/second order strongly hyperbolic evolution system\cite{oSgCjPmT02,gNoOoR04,hBoS04}. There are two ways of applying the theory discussed in this article to this problem. The first consists in finding a pseudodifferential first-order reduction of the evolution system which is non-characteristic. Such a reduction should be enough to impose constraint-preserving boundary conditions which are strongly stable, and for which the Kreiss theory can be applied. The second way is to exploit the non-uniqueness of time directions in general relativity and to write down a symmetric hyperbolic first-order reduction with respect to a suitable time evolution vector field, such that the boundary surface is non-characteristic and constraint-preserving boundary conditions can be specified in maximal dissipative form. We think that both  ways should, in the end, be equivalent but this needs further investigation. Regarding the BSSN system it turns out that it possesses zero speed fields which are intrinsic to the formulation\cite{gNoOoR04}, and therefore, it does not seem possible to apply the above methods unless one considers moving boundaries. Nevertheless, partial results\cite{dNoS10} have been recently obtained based on different methods.

\section*{Acknowledgments}

We thank H. Friedrich, H.-O. Kreiss and J. Winicour for numerous enlightening discussions on this subject. This work was supported in part by CONICET, SECYT-UNC, FONCYT and Grants No. CIC 4.19 to Universidad Michoacana and CONACyT 61173.

\bibliographystyle{unsrt}
\bibliography{refs_IBVP,reula}

\end{document}

%% file: prob.pstex_t
\begin{picture}(0,0)%
\includegraphics{prob.pstex}%
\end{picture}%
\setlength{\unitlength}{1450sp}%
\begingroup\makeatletter\ifx\SetFigFont\undefined%
\gdef\SetFigFont#1#2#3#4#5{%
  \reset@font\fontsize{#1}{#2pt}%
  \fontfamily{#3}\fontseries{#4}\fontshape{#5}%
  \selectfont}%
\fi\endgroup%
\begin{picture}(8864,4221)(2319,-5170)
\put(6616,-4246){\makebox(0,0)[lb]{\smash{{\SetFigFont{5}{6.0}{\rmdefault}{\mddefault}{\updefault}{\color[rgb]{0,0,0}$\Sigma_0$}%
}}}}
\put(9721,-1501){\makebox(0,0)[lb]{\smash{{\SetFigFont{5}{6.0}{\rmdefault}{\mddefault}{\updefault}{\color[rgb]{0,0,0}$\partial_t$}%
}}}}
\put(9946,-2941){\makebox(0,0)[lb]{\smash{{\SetFigFont{5}{6.0}{\rmdefault}{\mddefault}{\updefault}{\color[rgb]{0,0,0}$N^a$}%
}}}}
\put(9811,-3841){\makebox(0,0)[lb]{\smash{{\SetFigFont{5}{6.0}{\rmdefault}{\mddefault}{\updefault}{\color[rgb]{0,0,0}$\cal{T}$}%
}}}}
\put(8821,-1591){\makebox(0,0)[lb]{\smash{{\SetFigFont{5}{6.0}{\rmdefault}{\mddefault}{\updefault}{\color[rgb]{0,0,0}$T^a$}%
}}}}
\end{picture}%

%% file: zip_4.pstex_t
\begin{picture}(0,0)%
\includegraphics{zip_4.pstex}%
\end{picture}%
\setlength{\unitlength}{3025sp}%
\begingroup\makeatletter\ifx\SetFigFont\undefined%
\gdef\SetFigFont#1#2#3#4#5{%
  \reset@font\fontsize{#1}{#2pt}%
  \fontfamily{#3}\fontseries{#4}\fontshape{#5}%
  \selectfont}%
\fi\endgroup%
\begin{picture}(5376,3591)(691,-2823)
\put(4005,-2745){\makebox(0,0)[lb]{\smash{{\SetFigFont{9}{10.8}{\rmdefault}{\mddefault}{\updefault}{\color[rgb]{0,0,0}$\Sigma_0$}%
}}}}
\put(2303,165){\makebox(0,0)[lb]{\smash{{\SetFigFont{9}{10.8}{\rmdefault}{\mddefault}{\updefault}{\color[rgb]{0,0,0}{$\cal{T}$}}%
}}}}
\put(795,585){\makebox(0,0)[lb]{\smash{{\SetFigFont{9}{10.8}{\rmdefault}{\mddefault}{\updefault}{\color[rgb]{0,0,0}{$\tilde{\cal{T}}$}}%
}}}}
\put(1403,-487){\makebox(0,0)[lb]{\smash{{\SetFigFont{9}{10.8}{\rmdefault}{\mddefault}{\updefault}{\color[rgb]{0,0,0}$\psi$}%
}}}}
\put(2175,-1125){\makebox(0,0)[lb]{\smash{{\SetFigFont{9}{10.8}{\rmdefault}{\mddefault}{\updefault}{\color[rgb]{0,0,0}$K$}%
}}}}
\put(766,-1276){\makebox(0,0)[lb]{\smash{{\SetFigFont{9}{10.8}{\rmdefault}{\mddefault}{\updefault}{\color[rgb]{0,0,0}$\tilde{K}$}%
}}}}
\end{picture}%